\documentclass[11pt,twoside]{article}
\usepackage{asp2010}

\resetcounters

\begin{document}

\title{Statistical Parallax Analysis of SDSS M Dwarfs}

\author{Suzanne L. Hawley$^{1}$,
	John J. Bochanski$^{2,3}$
		 Andrew A. West$^{4}$
\affil{$^{1}$Astronomy Department, University of Washington,
   Box 351580, Seattle, WA  98195, email: slh@astro.washington.edu}
\affil{$^2$Astronomy and Astrophysics Department, Pennsylvania
  State University, 525 Davey Laboratory, University Park, PA 16802,  
email: jjb29@psu.edu}
\affil{$^3$Kavli Institute for Astrophysics and Space Research, Massachusetts Institute of Technology, Building 37, 77 Massachusetts Avenue, Cambridge, MA 02139}
\affil{$^4$Department of Astronomy, Boston University, 725 Commonwealth Avenue, Boston, MA 02215, email: aawest@bu.edu}}

\begin{abstract}
We report on the analysis of $\sim$ 22,000 M dwarfs
using a statistical parallax method.  This technique employs 
a maximum--likelihood formulation 
to simultaneously solve for the absolute magnitude, velocity 
ellipsoid parameters and reflex solar motion of a homogeneous
stellar sample, and has 
previously been applied to
Galactic RR Lyrae and Cepheid populations and to the 
Palomar/Michigan State University (PMSU) survey of nearby low-mass stars.   
We analyze subsamples of the most recent spectroscopic 
catalog of M dwarfs in the Sloan Digital Sky Survey (SDSS) 
to determine absolute magnitudes and kinematic properties as a function of 
spectral type, color, chromospheric activity and metallicity.
We find new, independent spectral type-absolute magnitude relations,
and color-absolute magnitude relations in the SDSS filters, and compare 
to those found from other methods.  Active stars have brighter absolute
magnitudes and lower metallicity stars have fainter absolute magnitudes
for stars of type M0-M4.  Our kinematic analysis confirms 
previous results for the solar motion and velocity dispersions, with more
distant stars possessing larger peculiar motions, and chromospherically 
active (younger) stars having smaller velocity dispersions 
than their inactive counterparts.  We find some evidence for 
systematic differences
in the mean $U$ and $W$ velocities of samples subdivided by color.
\end{abstract}

\section{Introduction}
M dwarfs are the most numerous stellar population in the 
Milky Way \citep{2010AJ....139.2679B}.  
Surveys such as the Sloan Digital Sky Survey 
\citep[SDSS;][]{2000AJ....120.1579Y}
and the Two Micron All Sky Survey
\citep[2MASS;][]{2006AJ....131.1163S} have led to photometric
and (in the case of SDSS) spectroscopic catalogs containing large
numbers of low mass stars.
The largest spectroscopic database
of M dwarfs \citep[][]{west10} compiles spectral types, colors, proper motions,
radial velocities and chromospheric activity estimates (as
traced by Balmer series emission) for more than 70,000 stars from
the most recent SDSS data release.

In order to make widespread use of these new measurements of 
the low--mass stellar population, with a focus on Galactic structure and kinematics, it is necessary to have good
distance estimates.  Since only very nearby M dwarfs have measured trigonometric
parallaxes, photometric and spectroscopic parallax relations
have typically been employed to obtain distances based on a star's color
or spectral type respectively \citep{2002AJ....123.3409H,
  2005PASP..117..706W, 2007ApJ...662..413K, 2008ApJ...673..864J}.  Because SDSS photometry
saturates at $m \sim$ 15, it is particularly difficult to anchor the photometric
parallax relations in the SDSS filters with measured trigonometric parallax stars.

The classical statistical parallax method is a way to determine 
the absolute magnitude of a homogeneous set of stars.  Statistical parallax
analysis seeks to determine the distance scale that provides the best match 
between the measured proper motions and radial velocities of a given 
stellar population, returning an estimate of
the average absolute magnitude of the population, and the kinematic properties
including the reflex solar motion and the velocity ellipsoid (velocity
dispersions along three principal axes).  Previous discussions of
the statistical parallax method can be found in
\cite{1971MNRAS.151..231C}, \cite{1983veas.book.....M} and \cite{1998ApJ...506..259P}.
Our particular formulation has been used to study 
RR Lyraes \citep
{1986ApJ...302..626H,
1986MNRAS.220..413S,
1996AJ....112.2110L,1998A&A...330..515F,1998ApJ...506..259P},
Cepheids \citep{1991ApJ...378..708W}, and the nearby low-mass
stars from the PMSU survey
\citep{1996AJ....112.2799H}.  

\section{Observations}\label{sec:obs}
Accurate photometry, proper motions and radial velocities are required for
statistical parallax analysis.  The data were obtained from the 
latest SDSS data release \citep[DR7; ][]{2009ApJS..182..543A} which
contains photometry over nearly 10,000 sq.\ deg.\ down to 
$r \sim$22 in five filters \citep[$ugriz$, ][]{1996AJ....111.1748F}.
When sky
conditions at Apache Point Observatory were not photometric, the SDSS
operated in a spectroscopic mode. 
Two fiber-fed spectrographs collected 640 spectra 
simultaneously with a typical total exposure time of $\sim$ 45 minutes.  These
medium--resolution ($R \sim 2,000$) spectra cover the entire optical bandpass
(3800 - 9200 \AA).  
Approximately 460,000 stellar spectra are available in the DR7 data
release, obtained both as targeted and serendipitous observations.  

The absolute astrometric precision of SDSS is $< 0.1^{\prime\prime}$ in each
coordinate \citep{2003AJ....125.1559P}.  Proper motions for SDSS
objects were found by matching to the USNO-B survey
\citep{2004AJ....127.3034M}.  The proper motions 
have a baseline of $\sim$ 50 years, and a typical precision
of $\sim$ 3 mas yr$^{-1}$ in right ascension and declination.
For a typical thin disk star with a transverse velocity of $\sim$ 10
km s$^{-1}$, this limit corresponds to a limiting distance of $\sim$
700 pc.  We therefore restricted the sample to stars closer than 700pc
as calculated from the color-magnitude relation of \cite{2010AJ....139.2679B},
to avoid biasing the solution by including only high velocity stars at
larger distances.

The \cite{west10} SDSS M dwarf sample was obtained by 
first selecting all stars in the DR7 spectroscopic catalog with colors 
$r-i > 0.42$ and $i-z > 0.24$, before 
correcting for Galactic reddening \citep{1998ApJ...500..525S}.  
Each spectrum was processed through the Hammer IDL
package \citep{2007AJ....134.2398C} which provides a spectral
type estimate and measures a number of spectral features, including
H$\alpha$ equivalent width and various TiO and CaH bandhead strengths.  
Each spectrum was also visually inspected and non-M dwarf contaminants 
were culled from the catalog.  The radial velocities were measured 
by cross--correlating each spectrum with the appropriate template from 
\cite{2007AJ....133..531B}, with a precision of $\sim$ 7 km s$^{-1}$.
To obtain the sample used in our statistical parallax analysis, 
we required precise photometry and accurate proper motions and radial
velocities.
See \cite{2010AJ....139.2679B} for information on photometric flag
cuts, and \cite{2010AJ....139.2566D} for our kinematic quality flags.  
After these quality cuts, our sample contained 40,963 
stars.  The 700pc distance cut further reduced the final statistical
parallax sample to 22,542 stars.

\section{Method and Sample Subdivision}\label{sec:method}
We employed the maximum--likelihood formulation of classical 
statistical parallax analysis as presented by \cite{1983veas.book.....M}.
Briefly, the velocity distribution of a homogeneous stellar 
population is modeled with nine kinematic parameters,
including the three orthogonal velocities of the reflex solar motion,
and the three directions and three dispersions of the
velocity ellipsoid, which describes the random and peculiar
velocities of the population.  
Two distance scale parameters allow the absolute magnitude and its
dispersion to vary; these are used
to transform the observed proper motions into transverse 
velocities.  There are thus eleven parameters used in the model.
In order to achieve well-determined solutions, we fixed the 
dispersion in the absolute magnitude to be $\sigma_M = 0.4$ which 
is the observed value for low--mass stars \citep{2010AJ....139.2679B}.

The observational data needed are
the position, apparent magnitude (corrected for Galactic extinction),
proper motions and radial velocity for each star in the
sample. 
We solve for the model parameters
simultaneously by maximizing the likelihood using
geometric simplex optimization 
\citep{optimizationsimplex_nelder_1965,1978Daniels}.  
Uncertainties in the parameters are
determined by numerical computation of the derivative
of each parameter individually, while keeping all other
parameters fixed.  The maximum--likelihood equations and 
simplex technique are described in detail in \cite{1986ApJ...302..626H}.

Our spectroscopic sample is much larger
than was available for previous statistical parallax studies, 
which typically contained observations of a few hundred or less objects.  
Thus, we were able to divide the
sample into smaller samples subdivided by color, spectral type,
magnetic activity (as traced by H$\alpha$), and metallicity (using the
$\zeta$ parameter of \citealp{2007ApJ...669.1235L}).
Each subsample of M dwarfs was analyzed using the following prescription.  
Ten runs were calculated for each dataset.  The initial absolute
magnitude estimates were derived from the $M_r, r-z$ color-magnitude
relation of \cite{2010AJ....139.2679B}.  The absolute
magnitude estimate was updated after each run, with the output of the
previous run being used as input for the next.  
For a given run, the simplex optimization iterated 5,000 times.
Thus, for each subsample, 50,000 iterations were computed.  This ensured
that the simplex operator had sufficient freedom to explore parameter
space and converged to the best (maximum likelihood) answer.  Typically,  
convergence was obtained after 25,000 iterations.  

\section{Results: Absolute Magnitudes}\label{sec:absmag}
Traditionally, colors or spectral types have been employed to
estimate absolute magnitude when trigonometric parallaxes were not
available.  These are referred to as photometric and
spectroscopic parallaxes, respectively.  In Figure 1
(left panel)
we show the spectroscopic parallax relation for
M dwarfs and compare it to previous studies
\citep{2002AJ....123.3409H, 2005PASP..117..706W, 2007ApJ...662..413K}.   
The open circles are nearby low--mass stars
with accurate trigonometric parallaxes, tabulated in
\cite{bochanskithesis}.  
Our results highlight the large dispersion in $M_r$ as a function of spectral
type.  The typical spread is $\sim$1 mag, increasing 
to $\sim$3 mag near type M4, much larger than the spread for a
given $r-z$ color bin (see Figure 1, right panel).  
This suggests that spectral type
is a poor tracer of $r$-band absolute magnitude for M dwarfs, 
since a single spectral type can encompass such a range of values.

Figure 1 (left panel) also displays disagreement between our
average statistical parallax result (solid black line,
filled circles) and the nearby star sample (open circles) at 
types later than about M4.  This is due to a systematic color difference
within a given spectral type bin between the two samples.  That is, 
the nearby star
sample is systematically redder than the SDSS sample at a given
spectral type, possibly due to the small number of nearby stars at
late types.  Thus, the SDSS stars have an absolute magnitude (at a
given spectral type) that is appropriate for the bluer, more luminous
stars of that type.  

Figure 1 (right panel) gives the statistical parallax $M_r, r-z$ relation, 
along with the results of previous studies \citep{2002AJ....123.3409H, 
2008AJ....135..785W, 2010AJ....139.2679B}. Overplotted with open
circles are the absolute magnitudes and colors from the nearby star
sample.  Note that the
dispersion in absolute magnitude as a function of color is
significantly smaller than that shown in Figure 1 (left panel)
indicating that the $r-z$ color is a much better tracer of $r$-band absolute
magnitude.  Furthermore, the discrepancy in mean color for a given
spectral type between the nearby star sample and the SDSS stars is not
evident.  We reiterate that colors, rather than spectral types, 
are preferred for absolute magnitude and distance estimation
for M dwarfs.

\begin{figure}[htbp]

\plottwo{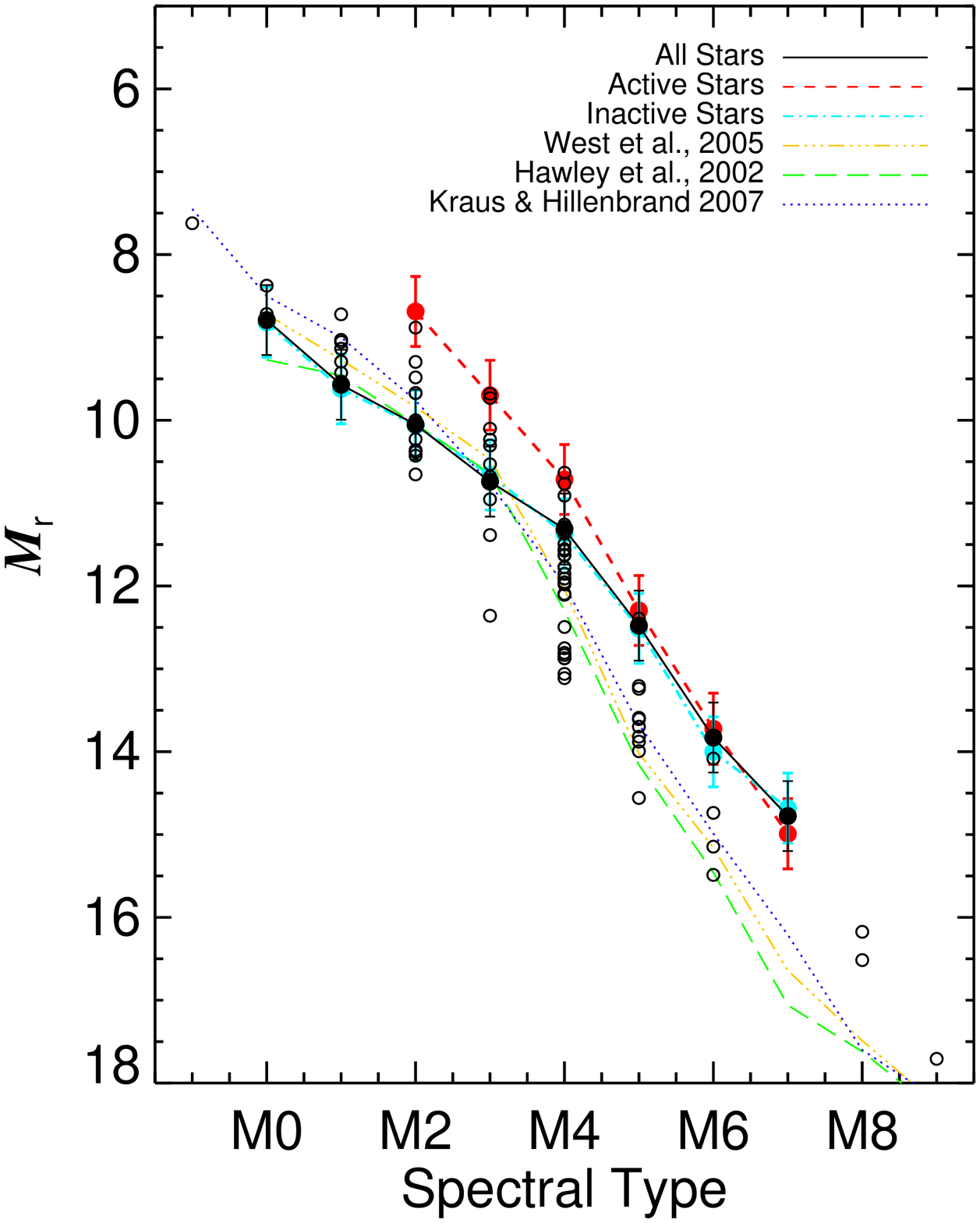}{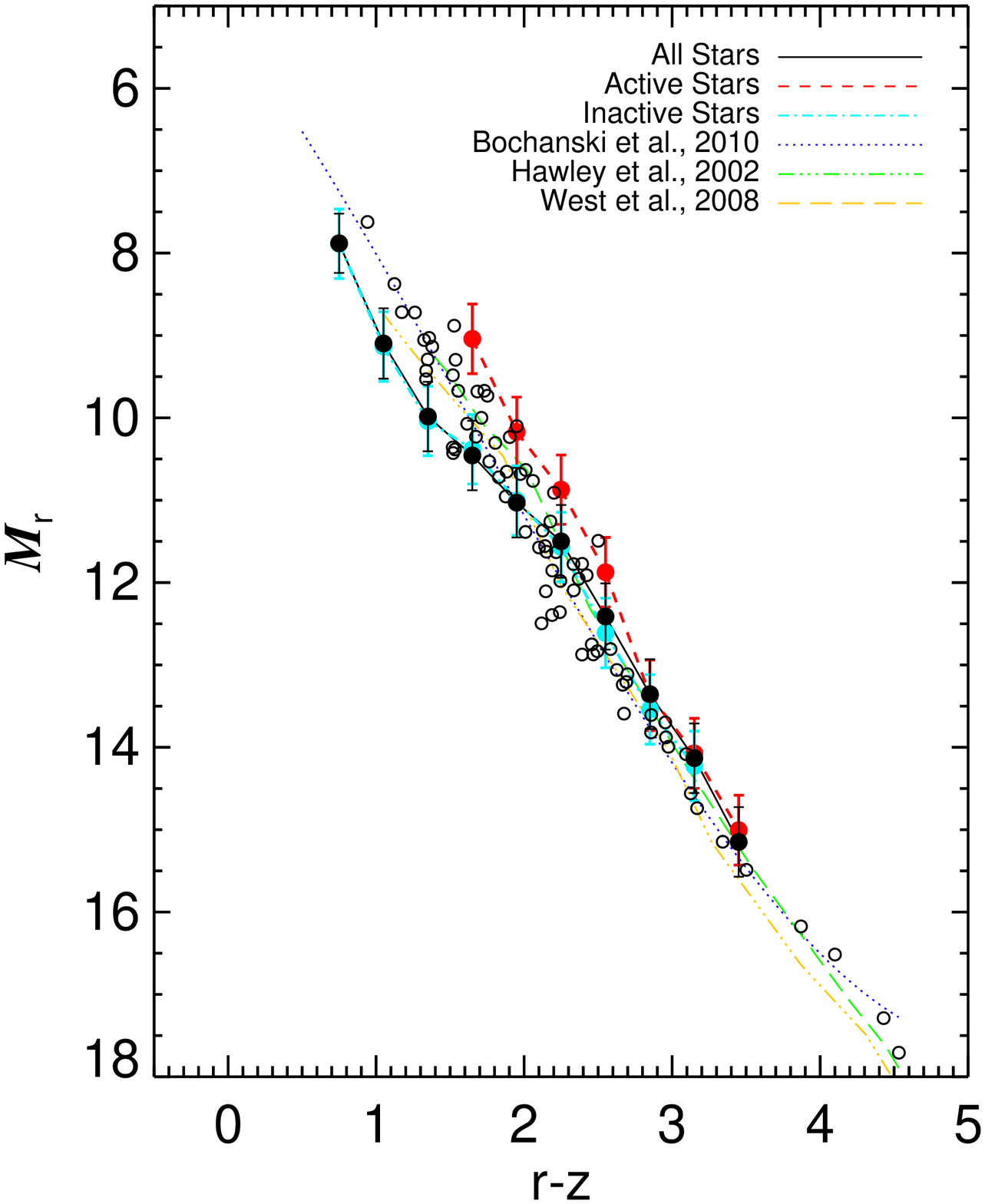} 
\caption{\textit{Left Panel} - $M_r$ vs. spectral type.  The symbols 
are described in the text.
  \textit{Right Panel} - $M_r$ vs. $r-z$, same symbols as in left panel.}
  \label{fig:mr_sp_rz}
\end{figure}

Due to the large spread in $M_r$ as a function of spectral type, we
concentrate on the color results shown in Figure 1 (right panel) to investigate
magnetic activity effects.
Our results are shown with the black, red and blue lines. 
There are a few notable trends.  First, the
bifurcated main sequence at blue colors (earlier spectral types) indicates that
active, early--type M dwarfs are about $\sim$ 1 mag brighter
in $M_r$ than inactive stars of the same color.  
A similar effect was observed in the PMSU sample, where active 
M dwarfs were brighter in $V$ than inactive stars (Figure 4 of 
\citealp{1996AJ....112.2799H}).
This difference could
be explained if the active stars have larger radii, as suggested by
recent observations of eclipsing binaries
\citep{2007ApJ...660..732L,2010ApJ...718..502M} and active, single stars \citep{2006ApJ...644..475B,2008A&A...478..507M}.
Direct
interferometric measurements of stellar radii will be necessary to
independently quantify the amount the radius changes for 
an individual star, since our
statistical analysis is performed on subsamples of stars that
span a small range of temperature.

At blue colors ($r-z < 2$), the inactive and total sample absolute
magnitudes fall below the mean locus of nearby trigonometric parallax
stars.  Since the
nearby stars are composed of a mix of active and inactive stars,
activity is likely not the only important effect on the luminosity.

We next investigated the effects of metallicity on $M_r$ and attempted
to isolate them from those traced by chromospheric activity.
Metallicity can alter a star's effective temperature and luminosity, 
manifesting as a change in absolute magnitude.
At a given color (or spectral type), stars with lower
metallicity typically exhibit fainter absolute magnitudes.
We examined the effects of metallicity 
using the $\zeta$ parameter, as defined by
\cite{2007ApJ...669.1235L}.  $\zeta$ is a metallicity proxy that can
be employed as a rough tracer of $[Fe/H]$, with $\zeta = 1$ corresponding
to solar metallicity and $\zeta = 0.4$ corresponding to $[Fe/H] \sim
-1$ \citep{2009PASP..121..117W}.  We note that the $[Fe/H], \zeta$
relation is only calibrated for early spectral types ($\sim$ M0-M3) 
and has significant uncertainty near solar metallicity.

In Figure 2, we plot the absolute magnitudes for two values of
$\zeta$ as a function of spectral type (left panel) and $r-z$ color
(right panel) for active and inactive stars.  
The uncertainty in each bin is $\sigma \approx 0.4$ mags which was fixed in the solution (see above).
Figure 2 demonstrates that for stars with
similar chromospheric properties, the lower metallicity ones have 
fainter absolute
magnitudes at the same color or spectral type, as expected.
Active stars at the
same metallicity are brighter than their inactive counterparts, but
both metallicity and activity are important for determining the 
luminosity of an individual star.
At a fixed $\zeta$,
the difference in $M_r$ is consistent with Figure 1,
with early--type active stars being $\sim 1$ mag brighter, and the
difference diminishing at later spectral types.  
The total SDSS sample is
overplotted in each panel with a solid black line.  The
early--type stars, which are seen at larger distances due to SDSS
magnitude limits, fall 
near the lower metallicity, inactive loci. The
later type stars, located closer to the Sun, are consistent with solar
metallicities and active stars.  This suggests that the low--mass dwarfs 
are tracing
a metallicity gradient similar to the one observed in SDSS
observations of higher--mass stars \citep{Ivezic08}, and also
explains why the SDSS sample falls below the locus of nearby stars
at early types in Figure 1.

We also note that the activity--metallicity loci appear to converge
near type M5 ($r-z \sim 2.8$).  This behavior is not well explained, but
may be linked to the transition between a partially and fully 
convective stellar
interior that occurs near that spectral type/color.
Perhaps this transition, which 
alters the efficiency of energy transport in the star, also regulates
the luminosity at the surface.

\begin{figure}[htbp]
 \plottwo{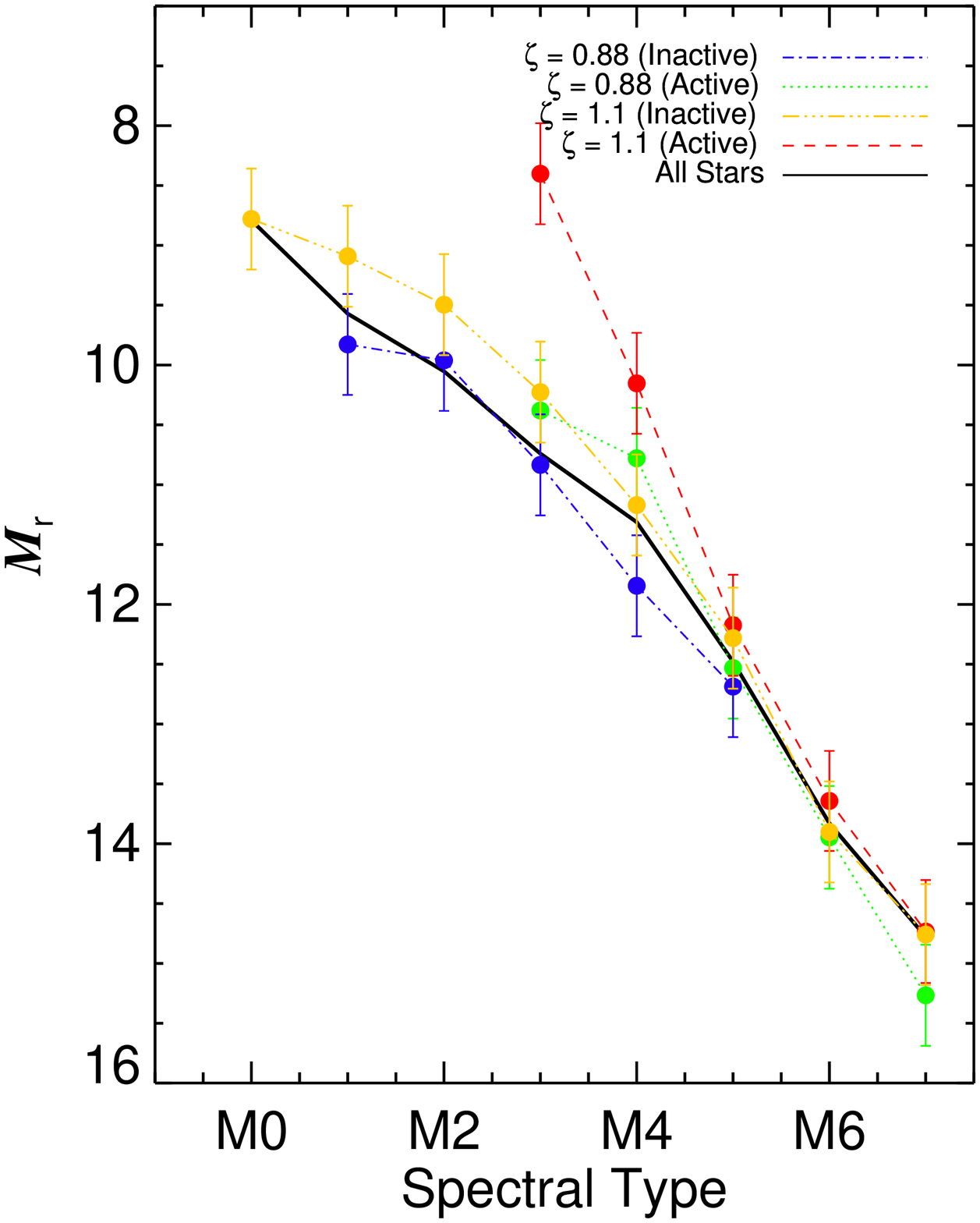}{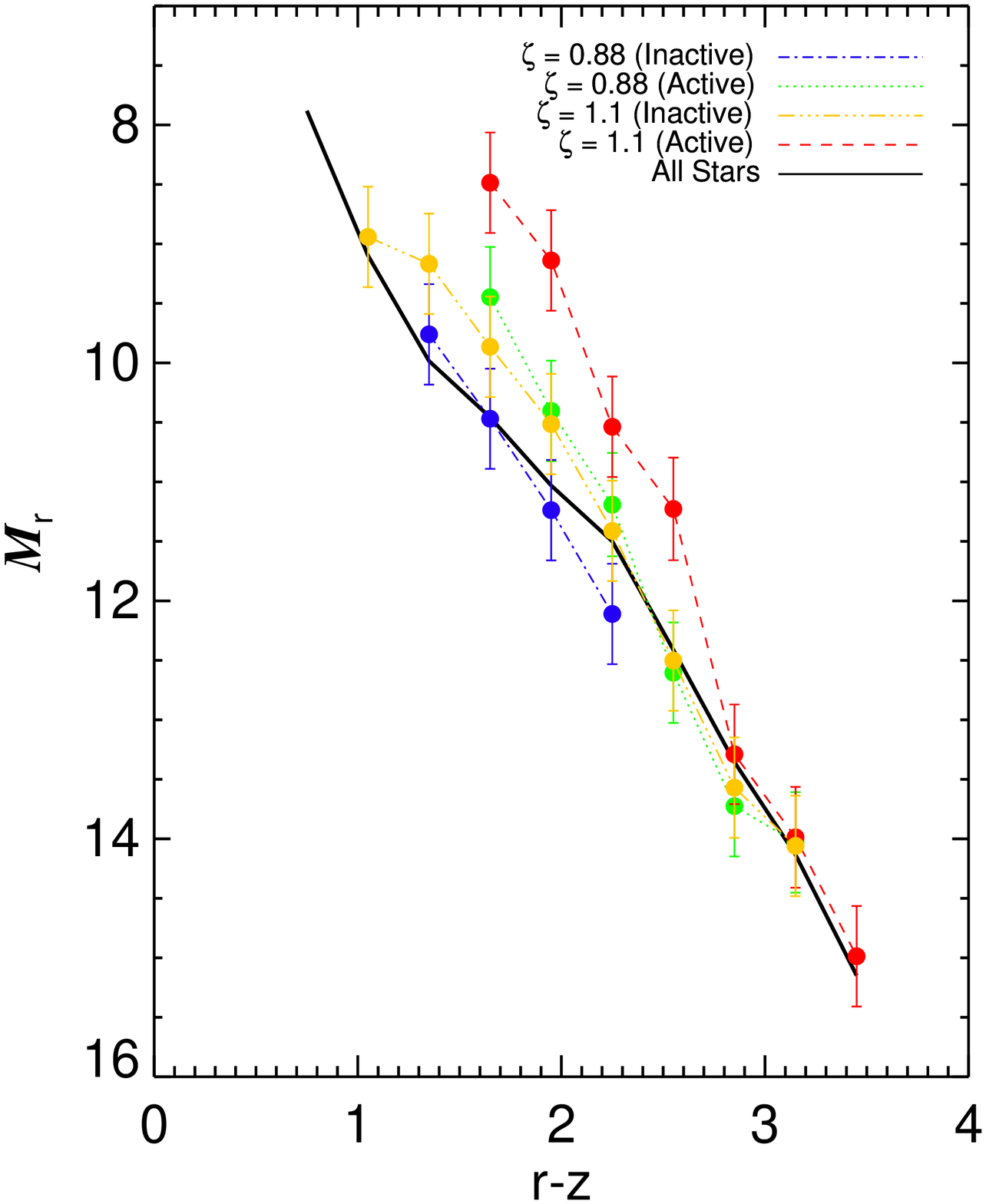}
\caption{\textit{Left Panel} - $M_r$ vs. spectral type for two values of
  $\zeta$, a metallicity proxy. 
   The total sample is overplotted in the solid black line. 
\textit{Right Panel} -  The $M_r$ vs. $r-z$ results show the same features
as in the spectral type panel.}
  \label{fig:mr_sp_rz_zeta}
\end{figure}

\section{Results: Kinematics}\label{sec:kinematics}
We measure kinematics in a $UVW$ coordinate system with 
$U$ increasing towards the Galactic
center, $V$ increasing in the direction of solar motion, and $W$
increasing vertically upward (as in \citealp{Dehnen98}).  

The statistical parallax method gives the 
reflex solar motion with respect to the mean velocity of the sample.
If samples possess
different mean velocities, this will be manifested as a change in the
solar reflex motions that are returned. 
In the $U$ and $W$ directions, the solar reflex motion reflects the 
Sun's peculiar motion, 
while in the $V$ direction it is a combination of the Sun's
peculiar motion and the asymmetric drift at the solar circle.
This increases the mean $V$ velocity of a sample of typical 
disk-age M dwarfs, 
giving a $V$ reflex motion that is larger than one measured from a 
population of young stars.  
Our results are shown in Figure 3 (left panel).
While the $U$ and $W$ velocities remain relatively
constant with spectral type, the $V$ reflex motion is significantly larger at
bluer colors.  These stars are observed at larger distances,
and also have larger velocity dispersions (as seen in Figure 3, right panel).
For $r-z > 2$, the $V$ velocity of the Sun remains relatively
constant at 20 km s$^{-1}$.  This value compares favorably to the
$V$ velocities previously measured for M dwarfs 
\citep{1996AJ....112.2799H,2009AJ....137.4149F}.
At later types, where the separation in age is
most extreme between the active and inactive populations,
the active stars in the 
sample show decreasing $V$ velocities, while the inactive stars
have increasing $V$ velocities as expected.

While the $U$ and $W$ velocities should
be constant with
color (or spectral type), we find that there is significant structure in both
distributions.  The $W$ velocity distribution in Figure 3 (left panel)
appears to peak near
$r-z \sim$ 2.3 (type M4).  This structure indicates the mean vertical
motions of the M dwarf subsamples are varying, 
with both bluer and redder M dwarfs 
having smaller mean $W$ velocities.
The $W$ velocities of active and inactive stars are not 
significantly different.  Meanwhile, the
$U$ velocity distribution shows a different behavior, 
exhibiting a rise toward later types which begins at a bluer
color in the active stars.  

\begin{figure}[htbp]
\plottwo{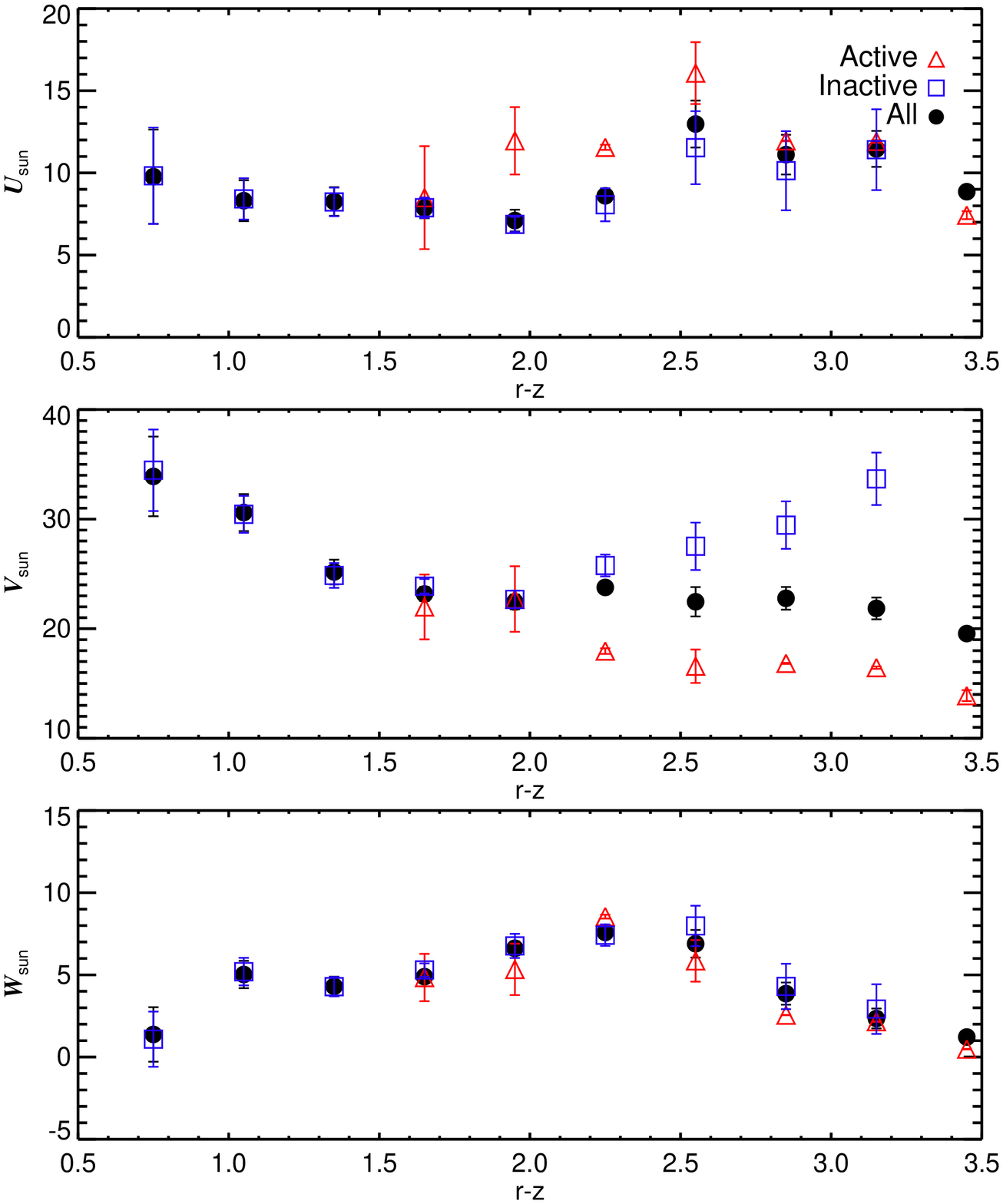}{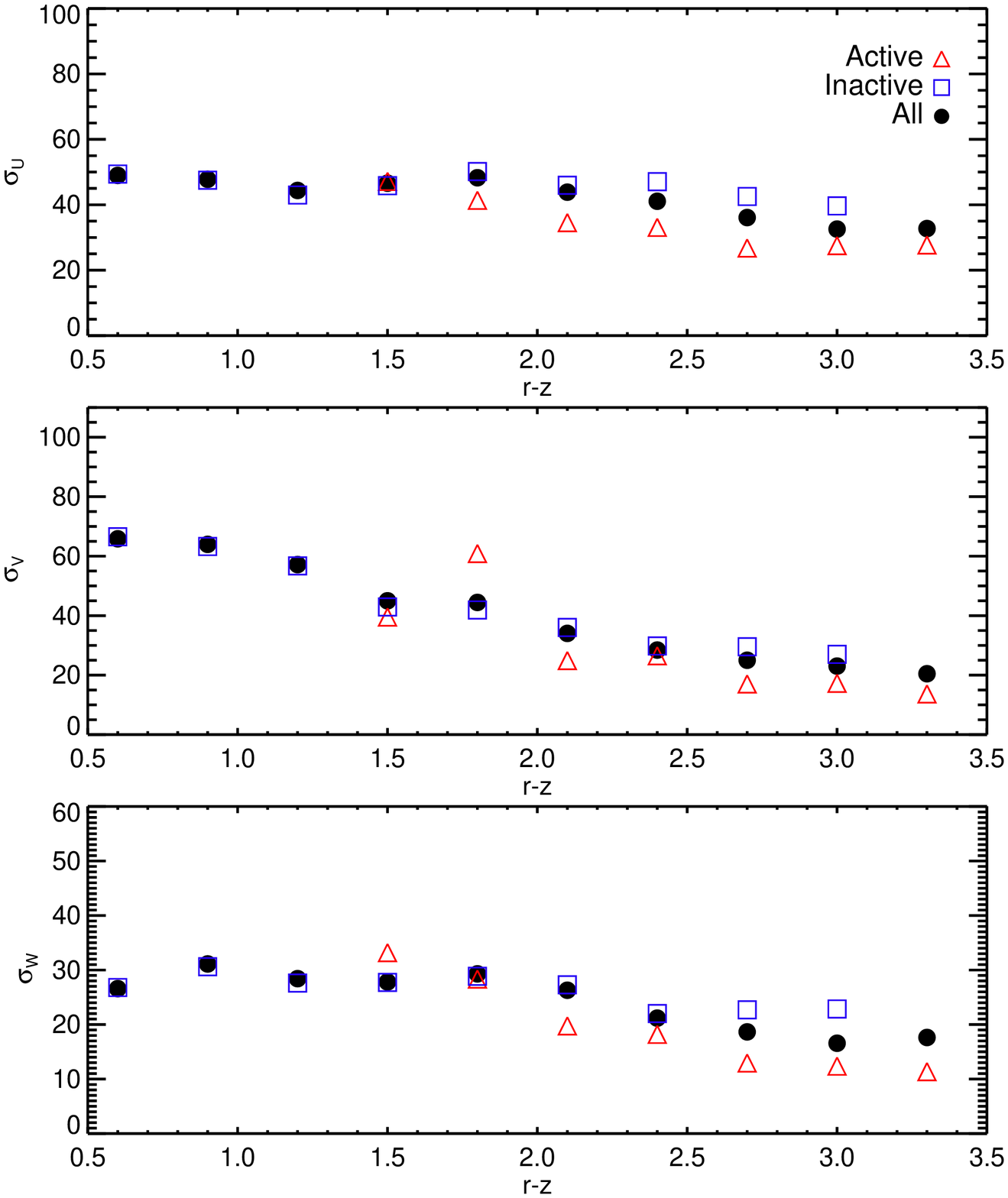} 
 \caption{\textit{Left Panel} - Solar peculiar motion as a function 
of $r-z$ color.  There is unexpected structure in the $U$ and $W$
distributions.  \textit{Right Panel} - $UVW$ velocity dispersions vs. 
color for active (red triangles), inactive (blue squares) and all stars 
(filled black circles).  Active stars possess smaller
    dispersions at red colors. }
  \label{fig:motions}
\end{figure}

The active M dwarfs are known to form a kinematically
colder population than their inactive counterparts \citep[e.g.,][]{1977A&A....60..263W}, as evidenced by smaller velocity dispersions.
This is usually interpreted
as an age effect, where younger M dwarfs have experienced less dynamical
heating.  Activity in low--mass stars has been shown to
depend on both spectral type and age; inactive, late--type M dwarfs are usually
much older than inactive early--type M dwarfs \citep{2008AJ....135..785W}.  
Figure 3 (right panel)
shows our velocity dispersion results.  As expected, the dispersions of
late--type (younger) active stars are smaller than for the late--type (older) 
inactive stars.  The overall decline 
in velocity dispersion toward redder colors reflects a similar effect,
since the early type stars are sampled further away, and are therefore
older and have higher dispersions than the closer, later--type stars.

We gratefully acknowledge the support of NSF grants AST 02-05875
and AST 06-07644 and NASA ADP grant NAG5-13111.   

Funding for the SDSS and SDSS-II has been provided by the Alfred P. Sloan 
Foundation, the Participating Institutions, the National Science Foundation, 
the U.S. Department of Energy, the National Aeronautics and Space 
Administration, the Japanese Monbukagakusho, the Max Planck Society, 
and the Higher Education Funding Council for England. 
The SDSS is managed by the Astrophysical Research Consortium for the 
Participating Institutions, which are listed at the SDSS 
Web Site http://www.sdss.org/.


\end{document}